\documentclass[aps,prl,reprint,superscriptaddress,longbibliography,floatfix]{revtex4-2}
\usepackage{amsmath,amssymb,graphicx,bm}
\usepackage[varg]{txfonts}
\usepackage[colorlinks=true,allcolors=blue]{hyperref}
\newcommand{\md}{\mathrm{d}}

\begin{document}
\title{Orientation-Dependent Universality of Spin-Conductance Fluctuations in d-Wave Altermagnets}

    \author{Yihan Zhang}
    \affiliation{Institute for Structure and Function \& Department of Physics \& Chongqing Key Laboratory for Strongly Coupled Physics, Chongqing University, Chongqing 400044, P. R. China}

    \author{Zheng Qin}
    \affiliation{Institute for Structure and Function \& Department of Physics \& Chongqing Key Laboratory for Strongly Coupled Physics, Chongqing University, Chongqing 400044, P. R. China}

    \author{Dong-Hui Xu}
    \affiliation{Institute for Structure and Function \& Department of Physics \& Chongqing Key Laboratory for Strongly Coupled Physics, Chongqing University, Chongqing 400044, P. R. China}
    \affiliation{Center of Quantum Materials and Devices, Chongqing University, Chongqing 400044, P. R. China}

    \author{Junjie Zeng}
    \email{jjzeng@cqu.edu.cn}
    \affiliation{Institute for Structure and Function \& Department of Physics \& Chongqing Key Laboratory for Strongly Coupled Physics, Chongqing University, Chongqing 400044, P. R. China}

    \author{Zhen Ning}
    \email{zhenning@cqu.edu.cn}
    \affiliation{Institute for Structure and Function \& Department of Physics \& Chongqing Key Laboratory for Strongly Coupled Physics, Chongqing University, Chongqing 400044, P. R. China}

    \author{Rui Wang}
    \email{rcwang@cqu.edu.cn}
    \affiliation{Institute for Structure and Function \& Department of Physics \& Chongqing Key Laboratory for Strongly Coupled Physics, Chongqing University, Chongqing 400044, P. R. China}
    \affiliation{Center of Quantum Materials and Devices, Chongqing University, Chongqing 400044, P. R. China}

    \date{\today}
\begin{abstract}
Universal conductance fluctuations reflect the symmetry and geometry of phase-coherent transport. However, it remains unclear whether crystal orientation relative to the device can tune fluctuation plateaus in spin-conserving altermagnets. Here, using quantum transport simulations of a two-dimensional $d$-wave altermagnet with scalar disorder, we show that crystal orientation tunes longitudinal and spin-Hall fluctuation  plateaus while each spin sector remains in the orthogonal symmetry class. At intermediate disorder, tuning the crystal orientation toward the device diagonal weakens longitudinal spin-conductance fluctuations and enhances spin-Hall fluctuations. Both responses exhibit well-defined plateaus at the diagonal orientation, with amplitudes that depend only weakly on disorder and altermagnetic strength over the ranges studied. A variance decomposition shows that, at this orientation, increasing disorder drives the longitudinal spin-channel covariance toward zero, while the transverse covariance shifts from enhancing to suppressing the spin-Hall variance. Our findings suggest that spin-conductance fluctuations can probe transport anisotropy in altermagnets.
\end{abstract}
\maketitle
\raggedbottom

\begin{figure*}[!tbp]
\centering
\includegraphics[width=\textwidth]{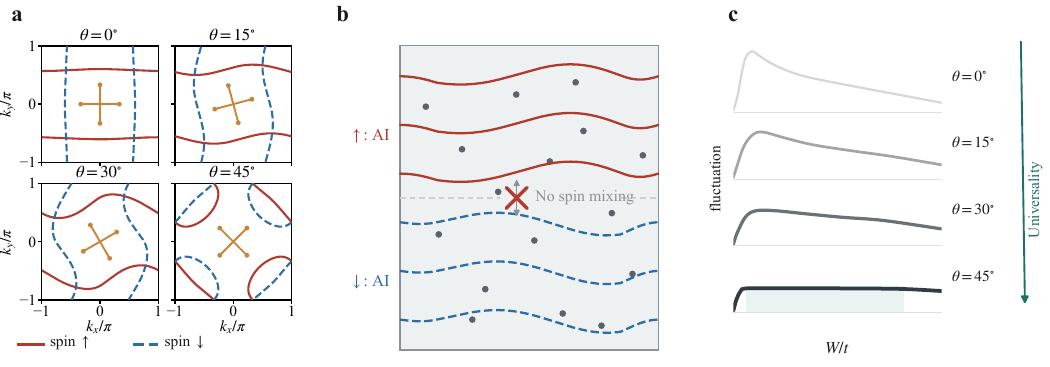}
\caption{Orientation-controlled fluctuation plateaus in a spin-conserving altermagnet. (a) Clean-band Fermi contours at orientation angles $\theta=0^\circ$, $15^\circ$, $30^\circ$, and $45^\circ$, calculated from Eq.~\eqref{eq:model} at Fermi energy $E_{\mathrm F}=t$ and altermagnetic strength $t_{\mathrm{AM}}=0.9t$, where $t$ is the nearest-neighbor hopping. Red solid and blue dashed curves denote spin up and down. Orange crosses indicate the orientation of the continuum $d$-wave splitting. The full-lattice contours change both orientation and shape. (b) Schematic interference paths in the two orthogonal (class AI) spin sectors of the same sample.
Gray scatterers represent the shared onsite potential $\{\varepsilon_i\}$.
The crossed vertical arrow indicates spin conservation. The shared disorder can still correlate the two channel conductances whose difference gives the spin conductance. (c) Qualitative illustration of plateau formation.
The downward arrow illustrates the approach to a flatter plateau toward $45^\circ$; its direction does not encode the angular trend of either amplitude.}
\label{fig:concept}
\end{figure*}

\emph{Introduction.---}Universal conductance fluctuations (UCF) reveal quantum interference through reproducible, sample-specific variations in the conductance of phase-coherent disordered conductors~\cite{Altshuler1985,Stone1985,LeeStone1985,LeeStoneFukuyama1987,Skocpol1986}.
Their characteristic amplitude depends on symmetry class, degeneracy, dimensionality, geometry, and boundary conditions~\cite{Beenakker1997RMP,Muttalib1987PRL,Jalabert1990PRL,Marcus1992PRL}.
Spin--orbit and spin-flip scattering can change the symmetry class, whereas Zeeman splitting can reduce the variance by lifting spin degeneracy~\cite{Chandrasekhar1990PRB,Takai1992PRB,Folk2001PRL}.
In spin--orbit-coupled wires, field orientation can also modify UCF through changes in the effective symmetry class~\cite{Scheid2009SST}.
Band anisotropy provides another route to modifying the fluctuation amplitude.
For example, gate-dependent UCF in $\mathrm{Cd}_3\mathrm{As}_2$ have been attributed to Fermi-surface anisotropy~\cite{Xiao2023NanoLett}. Likewise, lattice calculations link fluctuation amplitudes to anisotropic dispersion~\cite{Yang2025PRBAnisotropicUCF}.

In spintronics~\cite{Zutic2004}, conductance fluctuations can reveal information beyond the ensemble-averaged spin response.
For example, magnetic graphene nanoribbons can exhibit finite longitudinal spin conductance in individual samples even when the ensemble mean vanishes~\cite{Wimmer2008PRL,Bundesmann2013PRB}.
In transverse transport, the spin-Hall effect provides a probe of spin currents~\cite{Hirsch1999,Murakami2003,Sinova2004,Sinova2015}. Spin-Hall conductance fluctuations can form plateaus in spin--orbit-coupled models~\cite{Ren2006PRL,Qiao2011MPLB,Qiao2008PRL,Vasconcelos2016PRB,Qiao2007PRL,Santana2020PRB}.
The spin-Hall response also depends on disorder and on the geometry of the sample and leads~\cite{Mishchenko2004PRL,Nikolic2006PRB}.
For instance, geometric correlations can yield a finite mean spin conductance beyond the predictions of isotropic random-matrix theory~\cite{Adagideli2010PRL}.
A complementary approach is to examine the effects of band anisotropy in a spin-conserving system.

Altermagnets provide such a platform by combining compensated collinear order with anisotropic nonrelativistic spin splitting~\cite{Smejkal2022PRX,Smejkal2022Landscape,Jungwirth2026NatureSymmetry,Song2025NatRevMater,Fukaya2025JPCM,Cheong2025npjQM}.
In the collinear limit without spin--orbit coupling, the spin projection along the N\'eel axis is conserved and the anisotropic splitting allows spin-polarized currents~\cite{Gonzalez2021PRL,Jungwirth2026NatPhys}.
Spin-resolved photoemission~\cite{Jiang2025NatPhys} and local spin sensing~\cite{Wang2026Nature} probe the spin structure of the metallic $d$-wave altermagnet $\mathrm{KV}_2\mathrm{Se}_2\mathrm{O}$. A related layered candidate exhibits spin--valley locking~\cite{Zhang2025NatPhys}.
Complementary transport probes include electrical spin injection in compensated Mn$_5$Si$_3$~\cite{Mencos2026NatureSpinInjection} and measurements of exchange-related and spin--orbit contributions to spin--charge conversion in $\mathrm{RuO}_2$~\cite{Bai2023PRL,Liao2024PRL}.
Orientation-dependent conductivity, magnetoresistance, and anomalous Hall transport~\cite{Dou2025PRB,Zhang2025PRLAMR,Jin2024PRB}, together with geometry-dependent spin-Hall currents~\cite{Wei2025PRB}, connect band anisotropy to measurable transport responses.
Related exchange-driven transverse currents occur in anisotropic ferromagnets~\cite{Belashchenko2024PRB}. In altermagnetic systems, electrical control of spin transport and torque has also been explored~\cite{Zhou2026PRBControl,Zhang2025NatCommun,Fu2025npjQM}.
These results motivate examining whether orientation also affects coherent spin-current fluctuations.

Disorder theory for altermagnets has addressed weak-localization corrections to magnetoconductivity~\cite{Hijano2024PRResearch} and marginal metallic behavior under spin-mixing magnetic disorder~\cite{Li2026PRLDisorder}.
For spin-conductance fluctuations, a distinct issue arises under scalar disorder. The spin channels are decoupled and experience the same random potential.
Their conductances can therefore be correlated, in which case the covariance contributes to the spin-conductance variance.
Band anisotropy also makes the orientation relative to the contacts relevant within a fixed symmetry class.
How this orientation affects UCF in spin-conserving altermagnets remains open.
In particular, can it tune the longitudinal and spin-Hall fluctuation plateaus while each spin sector stays in the same symmetry class?

In this Letter, we demonstrate orientation-controlled spin-conductance fluctuation plateaus in a two-dimensional $d$-wave altermagnet with scalar disorder.
In our lattice model, orientation reshapes the spin-resolved Fermi contours relative to fixed device axes [Fig.~\ref{fig:concept}(a)], while both spin sectors remain in class AI.
The sectors experience the same random potential without spin mixing [Fig.~\ref{fig:concept}(b)].
We combine quantum transport simulations over disorder ensembles with Fermi-velocity analysis and an exact variance decomposition. Details are given in the Supplemental Material (SM)~\cite{SupplementalMaterial}.
Figure~\ref{fig:concept}(c) illustrates how the plateau profiles become flatter toward the diagonal orientation.
Between $15^\circ$ and $45^\circ$, the longitudinal amplitude decreases while the spin-Hall amplitude increases within their respective intermediate-disorder windows.
At $45^\circ$, both plateaus depend weakly on disorder and altermagnetic strength over the tested ranges.
The two amplitudes correlate oppositely with Fermi-velocity anisotropy. At the diagonal orientation, we also examine how the covariance contribution to the spin-conductance variance changes with disorder.
These results establish orientation as a control parameter for conserved-spin fluctuations within a fixed symmetry class.
Devices patterned along different crystal directions of metallic $d$-wave candidates such as $\mathrm{KV}_2\mathrm{Se}_2\mathrm O$ could probe this orientation dependence~\cite{Jiang2025NatPhys}.

\begin{figure}[!htbp]
\centering
\includegraphics[width=\columnwidth]{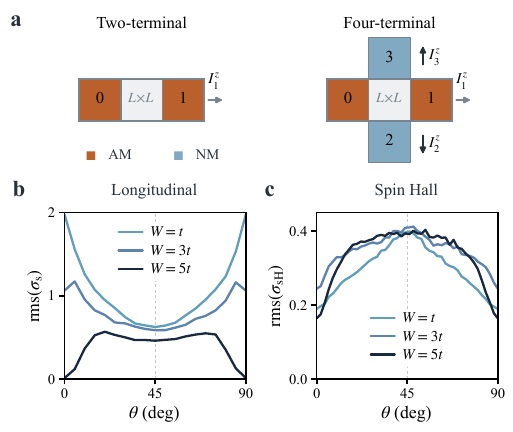}
\caption{Orientation dependence of spin-conductance fluctuations. (a) Two- and four-terminal devices with square disordered samples of side length $L$. Orange and blue contacts are altermagnetic (AM) and normal metal (NM), respectively. Arrows indicate the positive direction of the spin current $I_p^z$ into lead $p$. The upper and lower transverse contacts are leads 3 and 2, respectively. The transverse contacts are held at the equilibrium potential, as specified in the text. (b,c) Root-mean-square (rms) fluctuations of the longitudinal spin conductance $\sigma_{\mathrm s}$ and the spin-Hall conductance $\sigma_{\mathrm{sH}}$ versus orientation $\theta$. The curves show disorder strengths $W=t$, $3t$, and $5t$, from light to dark blue. Extended angular scans appear in Sec.~III of the SM~\cite{SupplementalMaterial}. All data use $E_{\mathrm F}=t$, $L=400$, $t_{\mathrm{AM}}=0.9t$, and 3000 disorder realizations per point. Conductance units are $e/(4\pi)$.}
\label{fig:orientation}
\end{figure}

\emph{Model and observables.---}We consider the $d$-wave altermagnetic lattice model of Refs.~\cite{Yi2025PRB,Sun2025TMR} with scalar onsite disorder. At each sample site $i$, the potential $\varepsilon_i$ is drawn independently and uniformly from $[-W/2,W/2]$ and acts equally on both spins. The width $W$ sets the disorder strength. The clean Bloch Hamiltonian is
\begin{equation}
\begin{aligned}
\mathcal H(\bm k)&=-2t(\cos k_x+\cos k_y)s_0+m_\theta(\bm k)s_z,\\
m_\theta(\bm k)&=2t_{\mathrm{AM}}\bigl[\cos(2\theta)(\cos k_x-\cos k_y)\\
&\hspace{24mm}-\sin(2\theta)\sin k_x\sin k_y\bigr].
\end{aligned}
\label{eq:model}
\end{equation}
Here $\bm k=(k_x,k_y)$ is crystal momentum. The angle $\theta$ parameterizes the orientation of the $d$-wave splitting relative to the longitudinal $x$ direction. The spin quantization axis remains fixed.
We set the lattice spacing $a=1$ and use the spin-independent nearest-neighbor hopping $t$ as the energy unit. Wave vectors and lengths are measured in units of $a^{-1}$ and $a$, respectively. The parameter $t_{\mathrm{AM}}$ sets the altermagnetic strength. In spin space, $s_0$ is the identity and $s_z$ is the Pauli matrix along the quantization axis. At fixed $\bm k$, the signed spin splitting $E_\uparrow-E_\downarrow$ equals $2m_\theta(\bm k)$.
Near $\Gamma$ ($\bm k=0$), $m_\theta$ reduces to a rotated $d$-wave form. On the full lattice, varying $\theta$ also changes the relative weights of the axial and mixed-direction harmonics, which alters the band shape.

Experimentally, relative orientation could be selected by patterning otherwise comparable devices along different in-plane crystal directions of a metallic $d$-wave altermagnet~\cite{Jiang2025NatPhys}.
To compare with such an experiment, one would rotate the device axes relative to a fixed material band structure.

For every $\theta$ in the model, the hopping is real and spin diagonal, and the onsite disorder is real and spin independent. Complex conjugation is therefore an antiunitary symmetry of each spin block and squares to $+1$.
Each block thus remains in class AI~\cite{AltlandZirnbauer1997}, although the altermagnetic order breaks physical spinful time-reversal symmetry.

We evaluate zero-temperature linear response at the Fermi energy $E_{\mathrm F}$ in two devices with a square disordered sample of side length $L$. The longitudinal device has two clean altermagnetic leads with the same band parameters as the sample. The four-terminal device for the transverse response adds two spin-conserving normal-metal leads [Fig.~\ref{fig:orientation}(a)].
Let $T_{ps,q}$ denote the total transmission from lead $q$ into lead $p$ within spin sector $s=\uparrow,\downarrow$. With $e>0$ the elementary charge, the channel conductances and spin-conductance differences are
\begin{equation}
\begin{aligned}
 Z_{\parallel,s}&=\frac{e}{4\pi}T_{1s,0},\\
 Z_{\perp,s}&=\frac{e}{16\pi}\bigl[(T_{2s,0}-T_{2s,1})-(T_{3s,0}-T_{3s,1})\bigr],\\
 Y_g&=Z_{g,\uparrow}-Z_{g,\downarrow},\qquad g=\parallel,\perp.
\end{aligned}
\label{eq:channel_weights}
\end{equation}
Here $g=\parallel,\perp$ denotes the longitudinal and transverse geometries, respectively. The applied voltage is $V=(\mu_0-\mu_1)/e$, where $\mu_p$ is the electrochemical potential of lead $p$. The current $I_p^z$ measures the flow of spin angular momentum along the quantization axis into lead $p$. Thus $Y_\parallel=\sigma_{\mathrm s}=I_1^z/V$ and $Y_\perp=\sigma_{\mathrm{sH}}=(I_2^z-I_3^z)/(2V)$. The difference between the channel conductances accounts for the opposite spin projections.

In the four-terminal device, $\mu_0=\mu_{\mathrm{eq}}+eV/2$, $\mu_1=\mu_{\mathrm{eq}}-eV/2$, and $\mu_2=\mu_3=\mu_{\mathrm{eq}}$, where $\mu_{\mathrm{eq}}$ is the equilibrium electrochemical potential. The transverse contacts thus collect spin current at fixed potential. The symmetric bias and the half-difference of the collected currents each contribute a factor of $1/2$ to the prefactor $e/(16\pi)$ in Eq.~\eqref{eq:channel_weights}.

Using the Landauer--B\"uttiker formalism~\cite{Landauer1970,Buttiker1986,Datta1995} implemented in Kwant~\cite{Groth2014}, we calculate $\mathrm{rms}(Y_g)=[\langle Y_g^2\rangle-\langle Y_g\rangle^2]^{1/2}$, where angle brackets denote the disorder average. Each transport point in the main figures uses 3000 disorder realizations, and both spin blocks see the same scalar potential in each realization. Lattice, transport, and ensemble definitions are detailed in Secs.~I, II, and IV of the SM~\cite{SupplementalMaterial}.

\begin{figure}[!htbp]
\centering
\includegraphics[width=\columnwidth]{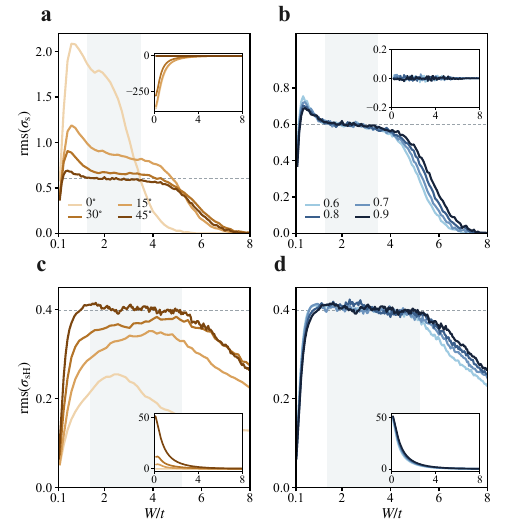}
\caption{Disorder dependence of longitudinal and spin-Hall conductance fluctuations. The upper and lower rows show the longitudinal and spin-Hall conductance fluctuation amplitudes versus normalized disorder strength $W/t$, respectively.
(a,c) Disorder scans at fixed $t_{\mathrm{AM}}=0.9t$ for $\theta=0^\circ, 15^\circ, 30^\circ,$ and $45^\circ$, from light to dark.
(b,d) Strength comparison at fixed $\theta=45^\circ$ for $t_{\mathrm{AM}}=0.6t, 0.7t, 0.8t$ and $0.9t$, from light to dark.
Each inset shows the corresponding disorder-averaged conductance versus $W/t$: $\langle\sigma_{\mathrm s}\rangle$ in (a,b) and $\langle\sigma_{\mathrm{sH}}\rangle$ in (c,d), with the same curve colors as the parent panel.
Shading marks approximate plateau regions; the averaging windows are specified in the text.
Dashed horizontal guides are drawn at 0.600 (longitudinal) and 0.399 (spin Hall); the window-averaged amplitudes are reported in the text.
All panels use $E_{\mathrm F}=t$, $L=400$, and 3000 disorder realizations per point. Fluctuation amplitudes and mean conductances are in units of $e/(4\pi)$.}
\label{fig:plateaus}
\end{figure}

\emph{Orientation-dependent universality.---}The angular scans in Fig.~\ref{fig:orientation}(b,c) distinguish the two responses. At $W=3t$, the longitudinal rms has a local minimum near $45^\circ$, while the spin-Hall rms has a broad maximum there. The extended angular scans, including the corresponding structures near $135^\circ$, are given in Sec.~III of the SM~\cite{SupplementalMaterial}. These variations occur without changing the class-AI symmetry of either spin block.

To see how these angular trends develop with disorder, we first compare scans at fixed $t_{\mathrm{AM}}=0.9t$ [Fig.~\ref{fig:plateaus}(a,c)]. The disorder profiles are strongly sloped at $0^\circ$ and flatten at larger angles. For the longitudinal response [Fig.~\ref{fig:plateaus}(a)], averaging over $1.6\leq W/t\leq3.2$ gives $0.868$, $0.662$, and $0.598$ at $15^\circ$, $30^\circ$, and $45^\circ$, respectively. The corresponding standard deviations across the sampled disorder points are $0.034$, $0.008$, and $0.008$. The larger residual slope at $15^\circ$ makes its plateau less well defined. In the transverse comparison window, $2.4\leq W/t\leq4.8$, the spin-Hall rms instead increases toward $45^\circ$ [Fig.~\ref{fig:plateaus}(c)]. These opposite trends describe the respective intermediate-disorder windows.

We next fix $\theta=45^\circ$ and vary the altermagnetic strength over $t_{\mathrm{AM}}/t=0.6$--$0.9$ [Fig.~\ref{fig:plateaus}(b,d)].
Both responses exhibit plateaus with weak dependence on disorder and altermagnetic strength within the sampled ranges.
Averaging over the four strengths and the windows above gives $0.595\pm0.009$ longitudinally and $0.399\pm0.006$ transversely. The quoted spreads are sample standard deviations of the rms amplitudes across the selected $(W,t_{\mathrm{AM}})$ points (see Sec.~IV of the SM~\cite{SupplementalMaterial}). The conductance unit is $e/(4\pi)$.
The shading highlights broader, visually plateau-like regions; the numerical averages use the explicitly stated windows.
The insets show the corresponding disorder-averaged conductances. Even where these averages are small, the sample-specific spin fluctuations remain finite.

The SM provides two further checks~\cite{SupplementalMaterial}. At $45^\circ$, energy scans show weak variation of both rms amplitudes around $E_{\mathrm F}=t$ (Sec.~III). Size scans support the stability of the spin-Hall plateau for $L=400$--$600$ at intermediate disorder (Sec.~V).

For comparison, spin--orbit-coupled models yield a spin-Hall rms near $0.18e/(4\pi)$ when the spin--orbit coupling is comparable to the hopping~\cite{Ren2006PRL}. A quantitative comparison requires consistent transverse-current definitions and an explicit treatment of channel correlations. Our conventions and covariance analysis are described in Secs.~II and VIII of the SM~\cite{SupplementalMaterial}.

\begin{figure}[!htbp]
\centering
\includegraphics[width=\columnwidth]{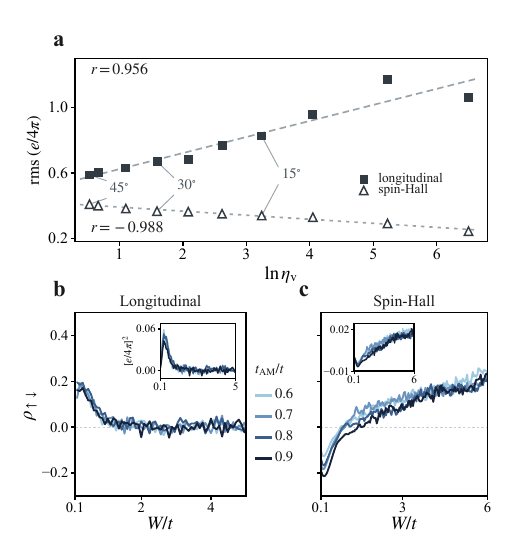}
\caption{Velocity anisotropy and spin-channel correlations.
(a) Longitudinal (filled squares) and spin-Hall (open triangles) fluctuation amplitudes at $W=3t$ versus $\ln\eta_{\mathrm v}$ for ten orientations from $0^\circ$ to $45^\circ$ in $5^\circ$ steps.
Here $\eta_{\mathrm v}$ is the largest-to-smallest eigenvalue ratio of the Fermi-velocity tensor. The Pearson coefficients for the correlations of the rms amplitudes with $\ln\eta_{\mathrm v}$ are $r=0.956$ (longitudinal) and $r=-0.988$ (spin Hall). Dashed lines are visual guides. Data in (a) use $t_{\mathrm{AM}}=0.9t$.
(b,c) Spin-channel correlation coefficient $\rho_g$ (labeled $\rho_{\uparrow\downarrow}$ in the figure) versus $W/t$ at $\theta=45^\circ$ for the longitudinal ($g=\parallel$) and spin-Hall ($g=\perp$) devices.
Insets show the corresponding covariances in units of $[e/(4\pi)]^2$. The shared legend for (b,c) gives $t_{\mathrm{AM}}/t=0.6$--$0.9$.
All panels use $E_{\mathrm F}=t$, $L=400$, and 3000 disorder realizations per point. Rms amplitudes are in units of $e/(4\pi)$.}
\label{fig:mechanism}
\end{figure}

\emph{Band anisotropy and spin-channel correlations.---}To relate the angular trends to the clean bands in Fig.~\ref{fig:concept}(a), we calculate the Fermi-velocity tensor $V^{(s)}$ and its eigenvalue ratio $\eta_{\mathrm v}$,
\begin{equation}
\begin{aligned}
 V_{ij}^{(s)}&=\frac{\displaystyle\int_{\mathrm{BZ}}\md^2k\,
 v_{i,s}v_{j,s}\delta_\gamma(E_s-E_{\mathrm F})}
 {\displaystyle\int_{\mathrm{BZ}}\md^2k\,\delta_\gamma(E_s-E_{\mathrm F})},\\
 \eta_{\mathrm v}&=\frac{\lambda_{\max}(V^{(s)})}{\lambda_{\min}(V^{(s)})}.
\end{aligned}
\label{eq:velocity_diagnostic}
\end{equation}
Here $i,j=x,y$, BZ denotes the Brillouin zone, and $\lambda_{\max}$ and $\lambda_{\min}$ are the largest and smallest tensor eigenvalues. The spin-sector dispersion $E_s(\bm k)$ follows from Eq.~\eqref{eq:model}. The component $v_{i,s}=\hbar^{-1}\partial E_s/\partial k_i$ is the group velocity along direction $i$, with $\hbar$ the reduced Planck constant. The kernel $\delta_\gamma(u)=\gamma/[\pi(u^2+\gamma^2)]$ is a normalized Lorentzian in the energy offset $u$, with half-width at half maximum $\gamma=0.032t$ (see Sec.~VII of the SM~\cite{SupplementalMaterial}). A $\pi/2$ rotation relates the two spin dispersions, so their velocity tensors have identical eigenvalues. Consequently, $\eta_{\mathrm v}$ is common to both blocks even though their principal axes differ.

At $E_{\mathrm F}=t$ and $t_{\mathrm{AM}}=0.9t$, $\eta_{\mathrm v}$ is $25.6$, $4.92$, and $1.70$ at $15^\circ$, $30^\circ$, and $45^\circ$, respectively. This decrease accompanies a lower longitudinal plateau and a higher spin-Hall plateau. To quantify the association at fixed disorder $W=3t$, we compare the rms amplitudes with $\ln\eta_{\mathrm v}$ across ten angles from $0^\circ$ to $45^\circ$. The Pearson correlation is positive for the longitudinal response ($r=0.956$) and negative for the spin-Hall response ($r=-0.988$) [Fig.~\ref{fig:mechanism}(a)].

In a diffusive description, band anisotropy enters the conductance fluctuations through the diffusion tensor. For a momentum-independent transport relaxation time $\tau_s$, the diffusion tensor is approximated by $D_{ij}^{(s)}\simeq\tau_s V_{ij}^{(s)}$. The corresponding diffusion modes obey the schematic eigenvalue equation~\cite{LeeStoneFukuyama1987}
\begin{equation}
 \left[-\partial_iD_{ij}^{(s)}\partial_j+\tau_\phi^{-1}\right]Q_\nu^{(s)}
 =\Lambda_\nu^{(s)}Q_\nu^{(s)},
\label{eq:diffusion_modes}
\end{equation}
where $Q_\nu^{(s)}(\bm r)$ is the $\nu$th diffusion mode in spin sector $s$ and $\Lambda_\nu^{(s)}$ is its relaxation rate. Here $\bm r=(r_x,r_y)$ and $\partial_i=\partial/\partial r_i$. Repeated spatial indices are summed, and the device sets the boundary conditions. The dephasing time $\tau_\phi$ is infinite in the coherent limit. With the contacts fixed, changes in the diffusion tensor can alter the mode spectrum within class AI. This gives a qualitative interpretation of the anisotropy dependence. A quantitative account of the spin-Hall variance would require the full four-terminal mode sum.

The angular variation of $\eta_{\mathrm v}$ reflects the full lattice dispersion. Near $\Gamma$, the principal quadratic coefficients $t\pm t_{\mathrm{AM}}$ are independent of $\theta$. The continuum ellipse rotates with $\theta$ at fixed eigenvalue ratio, so the variation at $E_{\mathrm F}=t$ requires the full lattice dispersion (see Secs.~I, VII, and IX of the SM~\cite{SupplementalMaterial}). At $\theta=0$ and $t_{\mathrm{AM}}=t$, the spin-up hopping along $x$ vanishes, illustrating the strong directional contrast approached at small angles.

Shared disorder can also correlate the two spin channels [Fig.~\ref{fig:concept}(b)]. Since $Y_g$ is their conductance difference in Eq.~\eqref{eq:channel_weights}, its variance $\mathrm{var}(Y_g)=\mathrm{rms}^2(Y_g)$ obeys
\begin{equation}
 \mathrm{var}(Y_g)=\mathrm{var}(Z_{g,\uparrow})+\mathrm{var}(Z_{g,\downarrow})-2\,\mathrm{cov}(Z_{g,\uparrow},Z_{g,\downarrow}).
\label{eq:variance}
\end{equation}
For disorder-dependent observables $A$ and $B$, $\mathrm{cov}(A,B)=\langle AB\rangle-\langle A\rangle\langle B\rangle$ and $\mathrm{var}(A)=\mathrm{cov}(A,A)$. When both channel variances are nonzero, the spin-channel correlation coefficient is $\rho_g=\mathrm{cov}(Z_{g,\uparrow},Z_{g,\downarrow})/[\mathrm{var}(Z_{g,\uparrow})\,\mathrm{var}(Z_{g,\downarrow})]^{1/2}$.

Relative to the sum of the channel variances, Eq.~\eqref{eq:variance} shows that positive covariance reduces the spin-conductance variance and negative covariance increases it. Both the channel variances and their covariance can depend on band structure and on how the device weights the transmissions in Eq.~\eqref{eq:channel_weights}.

At fixed $\theta=45^\circ$, the longitudinal channel correlation falls from about 0.2 at weak disorder to nearly zero when $W$ becomes comparable to $t$ [Fig.~\ref{fig:mechanism}(b)]. The longitudinal covariance also approaches zero at larger $W$, so the spin-conductance variance approaches the sum of the channel variances. In the transverse geometry, both the correlation and covariance change from negative to positive with increasing disorder [Fig.~\ref{fig:mechanism}(c)]. The covariance contribution therefore changes from enhancing to suppressing the spin-Hall variance.

\emph{Summary.---}Crystal orientation tunes longitudinal and spin-Hall fluctuation plateaus while each spin sector remains in class AI.
Both responses develop flatter disorder profiles toward $45^\circ$ and depend weakly on altermagnetic strength at that orientation over the ranges studied.
The angular trends correlate with Fermi-velocity anisotropy. At $45^\circ$, the channel covariance quantifies the correlation term in the spin-conductance variance.
An experimental test could compare geometrically matched devices patterned along different crystal directions of $\mathrm{KV}_2\mathrm{Se}_2\mathrm O$~\cite{Jiang2025NatPhys}.
Spin-sensitive detection could draw on lateral spin-valve measurements demonstrated for compensated Mn$_5$Si$_3$~\cite{Mencos2026NatureSpinInjection}. To probe coherent diffusive fluctuations with conserved spin, the devices should satisfy $\ell<L<\ell_\phi$ and have a spin-relaxation length exceeding $L$. Here $\ell$ and $\ell_\phi$ are the elastic mean free path and phase-coherence length, respectively (see Sec.~II of the SM~\cite{SupplementalMaterial}). Quantitative comparison with experiment would require material-specific transport calculations for the rotated devices, including spin--orbit coupling and dephasing.

\emph{Acknowledgments.---}This work was supported by the National Natural Science Foundation of China (Grants Nos. 12304191, 92365101, and 12347101) and the Fundamental Research Funds for the Central Universities (Grants Nos. 2023CDJXY-048 and 2024IAIS-ZX002).

\emph{Data availability.---}Supporting data are available from the corresponding authors upon request.

\nocite{Caroli1971,FisherLee1981,LopezSancho1984,LopezSancho1985}
\bibliographystyle{apsrev4-2}
\bibliography{references}
\end{document}


\setcounter{secnumdepth}{2}
\title{Supplemental Material for\\ Orientation-Dependent Universality of Spin-Conductance Fluctuations in $d$-Wave Altermagnets}
    
    \author{Yihan Zhang}
    \affiliation{Institute for Structure and Function \& Department of Physics \& Chongqing Key Laboratory for Strongly Coupled Physics, Chongqing University, Chongqing 400044, P. R. China}
    
    \author{Zheng Qin}
    \affiliation{Institute for Structure and Function \& Department of Physics \& Chongqing Key Laboratory for Strongly Coupled Physics, Chongqing University, Chongqing 400044, P. R. China}

    \author{Dong-Hui Xu}
    \affiliation{Institute for Structure and Function \& Department of Physics \& Chongqing Key Laboratory for Strongly Coupled Physics, Chongqing University, Chongqing 400044, P. R. China}
    \affiliation{Center of Quantum Materials and Devices, Chongqing University, Chongqing 400044, P. R. China}
    
    \author{Junjie Zeng}
    \email{jjzeng@cqu.edu.cn}
    \affiliation{Institute for Structure and Function \& Department of Physics \& Chongqing Key Laboratory for Strongly Coupled Physics, Chongqing University, Chongqing 400044, P. R. China}
    
    \author{Zhen Ning}
    \email{zhenning@cqu.edu.cn}
    \affiliation{Institute for Structure and Function \& Department of Physics \& Chongqing Key Laboratory for Strongly Coupled Physics, Chongqing University, Chongqing 400044, P. R. China}
    
    \author{Rui Wang}
    \email{rcwang@cqu.edu.cn}
    \affiliation{Institute for Structure and Function \& Department of Physics \& Chongqing Key Laboratory for Strongly Coupled Physics, Chongqing University, Chongqing 400044, P. R. China}
    \affiliation{Center of Quantum Materials and Devices, Chongqing University, Chongqing 400044, P. R. China}
    
    \date{\today}
\maketitle
\raggedbottom
\noindent This Supplemental Material provides the lattice and transport definitions,
ensemble statistics, extended parameter scans, and supporting analysis for the Letter.

\section{Lattice Hamiltonian, orientation, and symmetry}\label{sm:model}
\begin{figure}[t]
    	\centering
    	\includegraphics[width=0.58\textwidth]{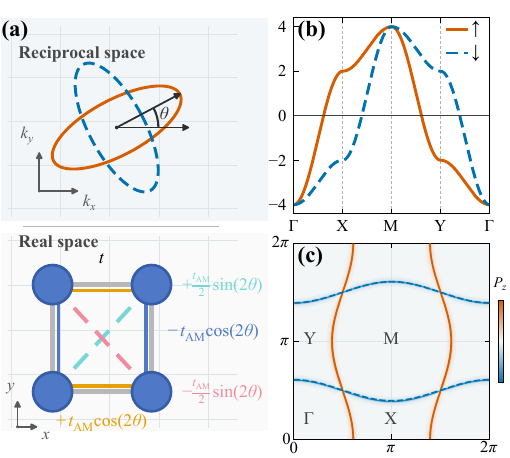}
    	\caption{\label{fig:sm_model}
    		Schematic illustration of the tight-binding model described by
    		Eqs.~\eqref{eq:hamiltonian_total} and~\eqref{eq:hamiltonian_am}, together with its spin-split electronic structure.
    		(a) The upper panel illustrates the orientation of the altermagnetic
    		anisotropy and the associated angle $\theta$ in reciprocal space. The lower
    		panel shows the corresponding real-space hopping pattern.
    		(b) Spin-resolved band structure along the high-symmetry path
    		$\Gamma$-$\mathrm{X}$-$\mathrm{M}$-$\mathrm{Y}$-$\Gamma$. The horizontal line marks the Fermi energy
    		$E_\mathrm{F}=0$. The orange solid and blue dashed curves represent the spin-up and
    		spin-down channels, respectively.
    		(c) Spin-resolved momentum-space spectral weight at $E_\mathrm{F}=0$, shown in the
    		equivalent Brillouin-zone representation
    		$0\leq k_x,k_y\leq 2\pi$. The color hue represents the spin polarization
    		$P_z=(A_{\uparrow}-A_{\downarrow})/(A_{\uparrow}+A_{\downarrow})$, while
    		the opacity reflects the total spectral weight
    		$A_{\uparrow}+A_{\downarrow}$.
    		For the numerical results in panels (b) and (c), we set
    		zero clean onsite potential, $t_{\mathrm{AM}}=0.5t$, and $\theta=0$.
    		The nearest-neighbor hopping $t$ is used as the energy unit, and a
    		Lorentzian broadening of $\gamma=0.032t$ is used in panel (c).
    	}
    \end{figure}

    We consider a square central region of size $L\times L$. Figure~\ref{fig:sm_model} shows the lattice model and its spin-resolved band structure. The two transport geometries are shown in Figs.~\ref{fig:sm_mean}(a,b). The two-terminal device measures the longitudinal spin conductance between two opposing altermagnetic leads. The four-terminal device adds two normal, spin-conserving transverse contacts to measure the spin-Hall conductance. The longitudinal leads match the altermagnetic sample, while the transverse contacts provide spin-conserving current detection.
    
    The disordered central region is described by a spinful tight-binding Hamiltonian
    \begin{equation}
        H = -t \sum_{\langle ij \rangle}\left(c_i^\dagger c_j+\mathrm{H.c.}\right) + \sum_i \varepsilon_i c_i^\dagger c_i + H_{\mathrm{AM}},
        \label{eq:hamiltonian_total}
    \end{equation}
    Here $c_i^\dagger = (c_{i\uparrow}^\dagger, c_{i\downarrow}^\dagger)$, and each nearest-neighbor bond $\langle ij\rangle$ is counted once. The hopping amplitude is $t$, and $\varepsilon_i$ is the onsite potential. In the disordered sample, $\varepsilon_i$ is drawn independently from a uniform distribution on $[-W/2,\,W/2]$, while in the ideal leads it is set to zero.
    
    Following the $d$-wave altermagnetic lattice model of Refs.~\cite{Yi2025PRB,Sun2025TMR}, we write the altermagnetic contribution as anisotropic spin-dependent hopping terms,
    \begin{align}
        H_{\mathrm{AM}} = t_{\mathrm{AM}} \sum_i \Bigl[
        &\cos(2\theta)\bigl(c_i^\dagger s_z c_{i+\hat{x}} - c_i^\dagger s_z c_{i+\hat{y}}\bigr) \notag\\*
        &+ \frac{\sin(2\theta)}{2}\bigl(c_i^\dagger s_z c_{i+\hat{x}+\hat{y}} - c_i^\dagger s_z c_{i+\hat{x}-\hat{y}}\bigr)
        + \mathrm{H.c.}
        \Bigr],
        \label{eq:hamiltonian_am}
    \end{align}
    Here $t_{\mathrm{AM}}$ sets the altermagnetic strength, and $s_z$ is the Pauli matrix in spin space. The angle $\theta$ specifies the crystalline anisotropy relative to the transport direction. Because only $s_z$ appears, the transport problem remains block diagonal in the spin basis used here. This exact spin conservation distinguishes the model from conventional spin--orbit-coupled spin-Hall models.
    
    In the clean limit, the Hamiltonian takes the compact momentum-space form
    \begin{equation}
        \mathcal H(\bm{k})=\varepsilon(\bm{k})s_0+m_{\mathrm{AM}}(\bm{k},\theta)s_z,
        \label{eq:kspace_hamiltonian}
    \end{equation}
    where
    \begin{equation}
        \varepsilon(\bm{k})=-2t(\cos k_x+\cos k_y),
        \label{eq:epsilon_k}
    \end{equation}
    and
    \begin{equation}
        \begin{split}
            m_{\mathrm{AM}}(\bm{k},\theta)=2t_{\mathrm{AM}}\bigl[&\cos(2\theta)(\cos k_x-\cos k_y)\\
            &-\sin(2\theta)\sin k_x\sin k_y\bigr].
        \end{split}
        \label{eq:am_splitting_k}
    \end{equation}
    
    Thus the two conserved spin sectors have dispersions
    \begin{equation}
        E_{\sigma}(\bm{k})=\varepsilon(\bm{k})+\sigma m_{\mathrm{AM}}(\bm{k},\theta),\qquad \sigma=\pm1.
        \label{eq:spin_resolved_dispersion}
    \end{equation}

    An instructive limiting case occurs at $\theta=0$, where the effective nearest-neighbor hoppings are $t_{x,\sigma}=t-\sigma t_{\mathrm{AM}}$ and $t_{y,\sigma}=t+\sigma t_{\mathrm{AM}}$. At $t_{\mathrm{AM}}=t$, the $\sigma=+1$ sector has $t_{x,+}=0$ and is therefore completely blocked for transport along $x$. This limit illustrates how directional hopping can suppress a transport channel without spin mixing.
    
    Near the $\Gamma$ point, the spin splitting becomes
    \begin{equation}
        m_{\mathrm{AM}}(\bm{k},\theta) \approx -t_{\mathrm{AM}}\cos(2\theta)(k_x^2-k_y^2)-2t_{\mathrm{AM}}\sin(2\theta)k_xk_y,
        \label{eq:am_splitting_continuum}
    \end{equation}
    which has the rotated $d$-wave form. 
    On the full lattice, varying $\theta$ changes the relative weights of the two lattice harmonics and can modify both the Fermi-surface shape and the strength of the transport anisotropy. 
    Thus, away from the continuum limit, this parameterization is not generally a rigid rotation of a fixed band structure.
    
    For Anderson-type onsite disorder, which is scalar and does not couple to spin, each conserved-spin block is real in the site basis and admits complex conjugation as an effective antiunitary symmetry with square $+1$. 
    Each block therefore belongs to class AI for every $\theta$~\cite{AltlandZirnbauer1997}. 
    This effective symmetry differs from physical spinful time reversal, which exchanges the spin sectors and is broken by the altermagnetic order.

In this model, orientation refers to the crystalline anisotropy relative to
the device; the spin quantization axis remains fixed. The anisotropy under
discussion is the spin-dependent band and transport anisotropy, not a
calculated magnetocrystalline anisotropy energy or a spin--orbit-driven
rotation of the N\'eel vector. These distinctions matter when translating
the lattice parameterization into a material-specific device.
\section{Transport definitions and numerical implementation}\label{sm:transport}
    We calculate transport using the nonequilibrium Green's-function method and the Landauer-B\"uttiker scattering formalism~\cite{Landauer1970,Buttiker1986,Datta1995,Groth2014}. The retarded Green's function of the open sample is
    \begin{equation}
        G^{\mathrm{r}}(E) = \bigl[E \mathrm{I} - H_{\mathrm{C}} - \sum_p \varSigma_p^{\mathrm{r}}(E)\bigr]^{-1},
        \label{eq:green_function}
    \end{equation}
    where $H_{\mathrm{C}}$ is the central-region Hamiltonian and $\varSigma_p^{\mathrm{r}}$ is the self-energy of terminal $p$. The latter follows from the surface Green's function of the corresponding semi-infinite lead~\cite{LopezSancho1984,LopezSancho1985}.
    Since spin is conserved, the scattering matrix is block diagonal in the spin-up and spin-down sectors. The two blocks are dynamically decoupled but sample the same disorder configuration and need not be statistically independent.
    
    The total spin-resolved transmission is obtained from the Caroli formula, equivalent to the scattering-matrix expression through the Fisher--Lee relation~\cite{Caroli1971,FisherLee1981,Datta1995}
    \begin{equation}
        T_{ps,q}(E) =
        \mathrm{Tr}\!\left[
        \varGamma_{p,s}(E) G_s^{\mathrm{r}}(E)
        \varGamma_{q,s}(E) G_s^{\mathrm{a}}(E)
        \right],
        \label{eq:transmission_caroli}
    \end{equation}
    where $ G_s^{\mathrm{a}} = (G_s^{\mathrm{r}})^{\dagger} $ and $ \varGamma_{p,s}(E) = \mi[\varSigma_{p,s}^{\mathrm{r}}(E) - \varSigma_{p,s}^{\mathrm{a}}(E)] $ is the linewidth matrix of terminal $ p $ in spin sector $ s $.
    
    We take $e>0$ and define the longitudinal bias by $V=(\mu_0-\mu_1)/e$, where $\mu_q$ is the electrochemical potential of lead $q$. Spin current is positive for positive $z$-angular momentum flowing from the sample into the receiving lead. In both geometries, channel conductances use spin-conductance units but exclude the sign of the spin angular momentum. That sign enters through the difference between the two channel contributions. The longitudinal channel conductances for $\alpha=\uparrow,\downarrow$ are therefore defined as
    \begin{equation}
        \sigma_{\alpha}=\frac{e}{4\pi}T_{1\alpha,0},
        \label{eq:spin_resolved_longitudinal}
    \end{equation}
    and the longitudinal spin conductance is
    \begin{equation}
        \sigma_{\mathrm{s}}=\sigma_{\uparrow}-\sigma_{\downarrow}
        =\frac{e}{4\pi}\left(T_{1\uparrow,0}-T_{1\downarrow,0}\right).
        \label{eq:spin_conductance_two_terminal}
    \end{equation}
    Here $\sigma_{\mathrm{s}}=I_1^z/V$ denotes the total two-terminal spin conductance, with $I_1^z$ the spin current entering lead 1. The minus sign accounts for the opposite spin angular momenta, as in the four-terminal definition below.

    In the four-terminal geometry, the longitudinal bias is symmetric: $\mu_0=\mu_{\mathrm{eq}}+eV/2$ and $\mu_1=\mu_{\mathrm{eq}}-eV/2$. Both transverse terminals are held at the equilibrium electrochemical potential, $\mu_2=\mu_3=\mu_{\mathrm{eq}}$. The antisymmetric spin-Hall conductance is $\sigma_{\mathrm{sH}}=(I_2^z-I_3^z)/(2V)$, where $I_p^z$ is the spin current entering transverse lead $p$. For each spin channel $s=\uparrow,\downarrow$, the transverse channel conductance follows the convention of Eq.~\eqref{eq:spin_resolved_longitudinal},
    \begin{equation}
        \sigma_{\mathrm{H},s} = \frac{e}{16\pi}\Bigl[(T_{2s,0} - T_{2s,1}) - (T_{3s,0} - T_{3s,1})\Bigr],
        \label{eq:spin_resolved_hall}
    \end{equation}
    where $T_{ps,q}$ is the total transmission from longitudinal lead $q=0,1$ into transverse lead $p=2,3$ within spin sector $s$. Spin conservation makes the incoming-spin label implicit in this notation. The factor $1/4$ relative to $e/(4\pi)$ combines the symmetric bias and the half-difference of the two transverse currents. The signs within each channel specify the spatial Hall combination. The opposite spin angular momenta enter through the difference defining the total spin-Hall conductance,
    \begin{equation}
        \begin{aligned}
            \sigma_{\mathrm{sH}} &= \sigma_{\mathrm{H},\uparrow} - \sigma_{\mathrm{H},\downarrow} \\
            &= \frac{e}{16\pi}\Bigl[
            (T_{2\uparrow,0} - T_{2\downarrow,0})
            - (T_{2\uparrow,1} - T_{2\downarrow,1}) \\
            &\qquad\qquad
            - (T_{3\uparrow,0} - T_{3\downarrow,0})
            + (T_{3\uparrow,1} - T_{3\downarrow,1})
            \Bigr].
        \end{aligned}
        \label{eq:spin_hall_total}
    \end{equation}
    Throughout, all channel conductances and total spin conductances are reported in units of $e/(4\pi)$. 
    
    Disorder averaging is performed at fixed $(E,W,t_{\mathrm{AM}},\theta,L)$. Details of the ensemble construction and numerical consistency checks are provided in Sec.~\ref{app:ensemble}. For an observable $X$, we use
    \begin{equation}
        \langle X \rangle = \frac{1}{N_{\mathrm{s}}}\sum_{n=1}^{N_{\mathrm{s}}} X_n,
    \end{equation}
    and define the root-mean-square (rms) fluctuation for either total spin response $Y=\sigma_\mathrm{s},\sigma_{\mathrm{sH}}$ as
    \begin{equation}
        \mathrm{rms}(Y) =
        \sqrt{\langle Y^2 \rangle - \langle Y \rangle^2}.
        \label{eq:rms_sh}
    \end{equation}
    To compare the two geometries at the spin-channel level, let $Z_{\alpha}=\sigma_{\alpha}$ for the longitudinal device and $Z_{\alpha}=\sigma_{\mathrm{H},\alpha}$ for the Hall device, with $\alpha=\uparrow,\downarrow$. In both cases, the measured spin conductance is $Y=Z_{\uparrow}-Z_{\downarrow}$. We evaluate
    \begin{equation}
        \mathrm{cov}(Z_{\uparrow},Z_{\downarrow}) =
        \langle Z_{\uparrow} Z_{\downarrow} \rangle
        - \langle Z_{\uparrow} \rangle \langle Z_{\downarrow} \rangle,
        \label{eq:covariance}
    \end{equation}
    When both channel variances are nonzero, the corresponding correlation coefficient is
    \begin{equation}
        \rho_{\uparrow\downarrow} =
        \frac{\mathrm{cov}(Z_{\uparrow},Z_{\downarrow})}
        {\sqrt{\mathrm{var}(Z_{\uparrow})\,\mathrm{var}(Z_{\downarrow})}}.
        \label{eq:correlation}
    \end{equation}
    Because both measured spin responses are differences between channel contributions, their variance obeys
    \begin{equation}
        \begin{aligned}
            \mathrm{var}(Y)
            &= \mathrm{var}(Z_{\uparrow}) + \mathrm{var}(Z_{\downarrow})
            - 2\,\mathrm{cov}(Z_{\uparrow},Z_{\downarrow}) \\
            &= \mathrm{var}(Z_{\uparrow}) + \mathrm{var}(Z_{\downarrow})
            - 2\,\rho_{\uparrow\downarrow}
            \sqrt{\mathrm{var}(Z_{\uparrow})\,\mathrm{var}(Z_{\downarrow})},
        \end{aligned}
        \label{eq:variance_decomposition}
    \end{equation}
    All variances and covariances have units of $[e/(4\pi)]^2$, while $\rho_{\uparrow\downarrow}$ is dimensionless.
    The clean limit $W=0$ contains a single deterministic configuration, so both variances and the covariance vanish and the correlation coefficient is undefined there; the displayed scans therefore start at $W=0.1t$, where the variances are already finite.
    
    Energies, disorder strengths, and altermagnetic strengths are measured in units of the nearest-neighbor hopping $t$. All numerical transport calculations are implemented using the Kwant package~\cite{Groth2014}.

\subsection{Length scales and the diffusive regime}\label{sm:lengths}

    Two length scales control the regime in which the plateaus form. The numerical model is fully coherent: no dephasing term is included, and because neither spin--orbit nor magnetic scattering is present, spin is conserved exactly. The phase-coherence length $\ell_\phi$ and the spin-relaxation length are thus infinite by construction, so the conditions $L<\ell_\phi$ and a spin-relaxation length exceeding $L$ hold identically in the calculation; they constrain a physical realization of the model rather than the simulation itself. The condition $\ell<L$, by contrast, depends on the disorder strength and is estimated below.

    The elastic mean free path can be estimated from the disorder-averaged channel conductance. At $E=t$ the leads support $N_{\rm c}=267$ transverse channels, the value of $T_{1s,0}$ in the clean limit. For a diffusive square of side $L$, $T_{1s,0}\simeq N_{\rm c}\ell/L$, so that $\ell/L\simeq T_{1s,0}/N_{\rm c}$. At $\theta=\pi/4$ and $t_{\mathrm{AM}}=0.9t$ this gives

    \begin{equation}
    \begin{array}{c|cccccc}
    W/t & 0.8 & 1.6 & 2.4 & 3.2 & 4.0 & 4.8\\ \hline
    T_{1s,0} & 52.3 & 15.3 & 6.46 & 3.10 & 1.47 & 0.60\\
    \ell/a & 78 & 23 & 9.7 & 4.6 & 2.2 & 0.9\\
    L/\ell & 5.1 & 17 & 41 & 87 & 1.8\times10^2 & 4.4\times10^2\\
    k_{\rm F}\ell & 1.6\times10^2 & 48 & 20 & 9.7 & 4.6 & 1.9
    \end{array}
    \label{eq:mean_free_path}
    \end{equation}

    with $k_{\rm F}\simeq2.1$ obtained from $N_{\rm c}=k_{\rm F}L/\pi$. Across the longitudinal comparison window $1.6\leq W/t\leq3.2$ the mean free path therefore falls from about $23$ to $5$ lattice constants, $L/\ell$ rises from $17$ to $87$, and $k_{\rm F}\ell$ remains between $48$ and $10$; transport is diffusive with many open channels, and $\ell\ll L$ holds. At $W=4.8t$, the upper end of the spin-Hall comparison window, $\ell$ has dropped to about one lattice constant and $k_{\rm F}\ell\simeq2$; the spin-Hall plateau therefore extends to the edge of the diffusive regime, and the estimate itself becomes unreliable beyond it, consistently with the strong-disorder suppression of both amplitudes seen at $W\gtrsim6t$.
    
\section{Average responses and full energy, angle, and disorder scans}\label{sm:extended}
    \begin{figure}[!tbp]
        \centering
        \includegraphics[width=0.64\textwidth]{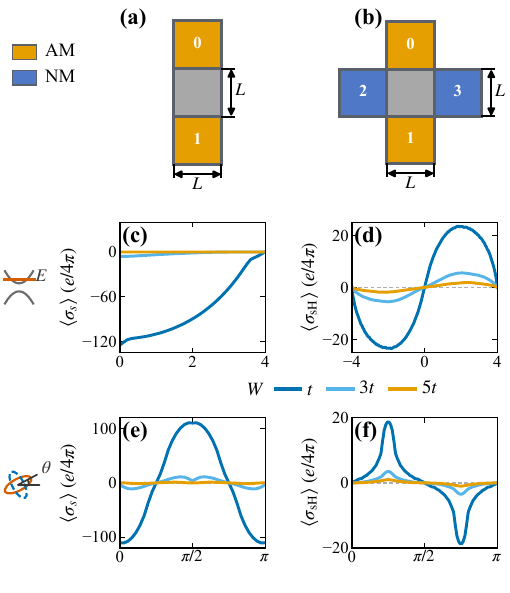}
        \caption{
        	Two- and four-terminal transport geometries and their corresponding
        	disorder-averaged responses.
        	(a,b) Two- and four-terminal geometries, respectively. The orange and
        	blue regions denote altermagnetic (AM) and normal-metal (NM) leads,
        	while the gray square represents the disordered scattering region of
        	size $L\times L$. The numbers label the lead indices.
        	(c,e) Disorder-averaged spin conductance
        	$\langle\sigma_{\mathrm{s}}\rangle$ for the two-terminal geometry shown in panel
        	(a), as functions of the Fermi energy $E$ and the altermagnetic
        	orientation angle $\theta$, respectively.
        	(d,f) Corresponding spin-Hall conductance
        	$\langle\sigma_{\mathrm{sH}}\rangle$ for the four-terminal geometry
        	shown in panel (b).
        	For the energy scans, $\theta=0$ in panel (c) and $\theta=\pi/4$ in
        	panel (d); for the angular scans in panels (e,f), $E=t$.
        	The curves represent disorder strengths $W=t$, $3t$, and $5t$.
        	All calculations use $L=400$ and $t_{\mathrm{AM}}=0.9t$, with the
        	nearest-neighbor hopping $t$ taken as the energy unit.
        	Each data point is averaged over at least $1000$ disorder realizations.
        }
        \label{fig:sm_mean}
    \end{figure}
    
    We first examine the disorder-averaged transport response in the two geometries introduced above. 
    Figures~\ref{fig:sm_mean}(c) and~\ref{fig:sm_mean}(d) show the energy dependence of the averaged longitudinal spin conductance at $\theta=0$ and the spin-Hall conductance at $\theta=\pi/4$, respectively. 
    In both geometries, increasing disorder rapidly suppresses the magnitude of the disorder-averaged response. 
    The four insets in Fig.~3 of the Letter show the corresponding disorder dependence for its angle and strength comparisons. We compare these mean responses with the fluctuation amplitudes below.
    
\subsection{Angular dependence of the average response}
    
    Figures~\ref{fig:sm_mean}(e,f) show the orientation dependence of the mean conductances. The two-terminal spin conductance is extremal close to $\theta=\pi/2$ and changes sign near $\theta=\pi/4$ and $3\pi/4$. The four-terminal spin-Hall signal is enhanced near $\theta\approx\pi/4$, reverses sign around $\theta=\pi/2$, and exhibits the opposite extremum near $3\pi/4$. Increasing disorder reduces both signal magnitudes while preserving their overall angular dependence in the displayed scans. 
    We next examine fluctuations about these disorder-averaged conductances.
    
    \begin{figure*}[!t]
        \centering
        \includegraphics[width=\textwidth]{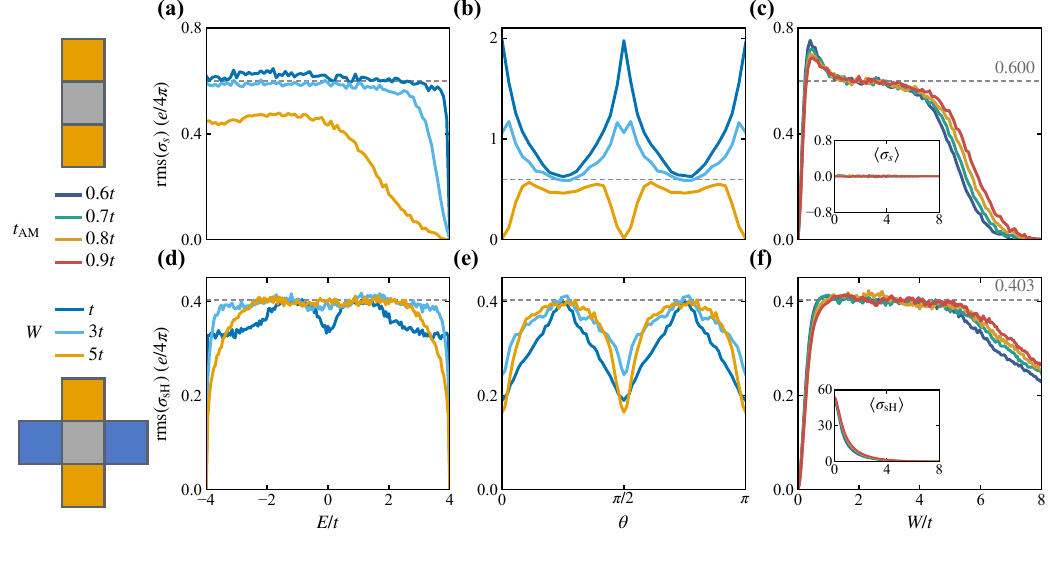}
        \caption{
        	Mesoscopic conductance fluctuations in the two- and four-terminal
        	geometries at $L=400$. The device schematics on the left identify the
        	geometry associated with each row.
        	(a--c) Spin-conductance fluctuation $\mathrm{rms}(\sigma_{\mathrm{s}})$ in the
        	two-terminal geometry.
        	(d--f) Spin-Hall conductance fluctuation
        	$\mathrm{rms}(\sigma_{\mathrm{sH}})$ in the four-terminal geometry.
        	Panels (a,d), (b,e), and (c,f) show the dependence on the Fermi energy
        	$E$, the altermagnetic orientation angle $\theta$, and the disorder
        	strength $W$, respectively.
        	For the energy and angular scans in panels (a,b,d,e),
        	$t_{\mathrm{AM}}=0.9t$ and $W=t$, $3t$, and $5t$.
        	The energy scans are calculated at $\theta=\pi/4$, while the angular
        	scans are calculated at $E=t$.
        	For the disorder scans in panels (c,f), $E=t$ and $\theta=\pi/4$, with
        	$t_{\mathrm{AM}}=0.6t$, $0.7t$, $0.8t$, and $0.9t$.
        	The gray dashed lines mark the reference fluctuation amplitudes
        	$0.600\,e/(4\pi)$ and $0.403\,e/(4\pi)$ for the two- and four-terminal
        	geometries, respectively.
        	The insets in panels (c,f) show the corresponding disorder-averaged
        	conductances $\langle\sigma_{\mathrm{s}}\rangle$ and
        	$\langle\sigma_{\mathrm{sH}}\rangle$.
        	The nearest-neighbor hopping $t$ is used as the energy unit, and each
        	data point is averaged over at least $1000$ disorder realizations.
        }
        \label{fig:sm_fluctuations}
    \end{figure*}
    The energy scans in Figs.~\ref{fig:sm_fluctuations}(a,d) are performed at $\theta=\pi/4$. 
    At weak and intermediate disorder, the longitudinal rms forms a broad plateau near $0.60\,e/(4\pi)$ in the central energy range. Its amplitude decreases as the Fermi energy approaches the band edges. 
    The spin-Hall rms remains near $0.40\,e/(4\pi)$ over much of the central energy window. 
    Although the detailed structures differ, both observables display broad energy intervals in which their amplitudes depend only weakly on energy. We consequently choose $E=t$ for the orientation and disorder scans.

\section{Ensemble construction and consistency checks}\label{app:ensemble}

    At each disordered transport point, the reported rms is the standard deviation over $N_{\mathrm{s}}\geq1000$ realizations, evaluated with normalization $1/N_{\mathrm{s}}$. When combining independently generated batches, we first concatenate the conductances from all realizations. We then recompute all averages, variances, covariances, and correlation coefficients from the combined ensemble. Statistics from separate batches are therefore not treated as individual disorder realizations. The clean limit $W=0$ contains a single deterministic configuration and is consequently assigned zero disorder-induced fluctuation.

    Both spin channels are evaluated using the same disorder realization. Before evaluating ensemble moments, we check $\sigma_{\mathrm{s}}=\sigma_{\uparrow}-\sigma_{\downarrow}$ and $\sigma_{\mathrm{sH}}=\sigma_{\mathrm{H},\uparrow}-\sigma_{\mathrm{H},\downarrow}$ for every realization in the two geometries. Here $\sigma_{\alpha}=(e/(4\pi))T_{1\alpha,0}$ for $\alpha=\uparrow,\downarrow$, and all channel contributions and total spin conductances are expressed in units of $e/(4\pi)$, as in the Letter. For the merged four-terminal datasets, the largest absolute residual in the latter identity is below $1.1\times10^{-9}$ in these conductance units.

    In both geometries, channel conductances exclude the spin-angular-momentum sign, which enters through the up-minus-down difference. In particular, $\sigma_{\mathrm{H},\downarrow}$ includes the spatial signs of the transverse Hall combination but no additional minus sign for the down-spin angular momentum. The channel covariances and correlation coefficients use this same convention.

    To estimate a representative plateau amplitude, we assign equal weight to each plotted parameter point within the stated window. If $R_\alpha=\mathrm{rms}(Y)_\alpha$ denotes the ensemble fluctuation at the parameter point $\alpha=(W,t_{\mathrm{AM}})$, we define
    \begin{equation}
        \begin{gathered}
        \overline{R}=\frac{1}{N_{\mathrm{p}}}\sum_{\alpha=1}^{N_{\mathrm{p}}}R_\alpha,\\
        \Delta_{\mathrm{pl}}=\left[\frac{1}{N_{\mathrm{p}}-1}
        \sum_{\alpha=1}^{N_{\mathrm{p}}}(R_\alpha-\overline{R})^2\right]^{1/2}.
        \end{gathered}
        \label{eq:plateau_extraction}
    \end{equation}
    Here $\Delta_{\mathrm{pl}}$ denotes the spread of the rms estimates across the selected parameter points. It includes both residual parameter dependence and finite-ensemble variability. It is neither the standard error of an individual rms estimate nor the uncertainty of a fitted universal constant.
    At $\theta=\pi/4$, averaging the four $t_{\mathrm{AM}}=0.6t$--$0.9t$ curves over $1.6\leq W/t\leq3.2$ for the longitudinal response and $2.4\leq W/t\leq4.8$ for the spin-Hall response gives
    \begin{equation}
        \overline{R}_{\mathrm{s}}=0.595\pm0.009\,\frac{e}{4\pi},
        \qquad
        \overline{R}_{\mathrm{sH}}=0.399\pm0.006\,\frac{e}{4\pi},
        \label{eq:plateau_values}
    \end{equation}
    The quoted $\pm$ values denote $\Delta_{\mathrm{pl}}$.
    These numerical averages use the explicitly stated windows, irrespective of the broader shading in Fig.~3 of the Letter.
    The latter highlights approximate plateau regions, $1.3\leq W/t\leq3.5$ longitudinally and $1.4\leq W/t\leq5.2$ transversely.
    The dashed lines at $0.600\,e/(4\pi)$ and $0.403\,e/(4\pi)$ in Fig.~\ref{fig:sm_fluctuations} serve as nearby visual guides.
    In Fig.~3 of the Letter, the dashed lines are visual guides at $0.600\,e/(4\pi)$ longitudinally and $0.399\,e/(4\pi)$ transversely. The corresponding window averages are given in Eq.~\eqref{eq:plateau_values}.

    \section{Robustness against system size and altermagnetic strength}\label{app:robustness}

\begin{figure*}[t]
        \centering
        \includegraphics[width=0.85\textwidth]{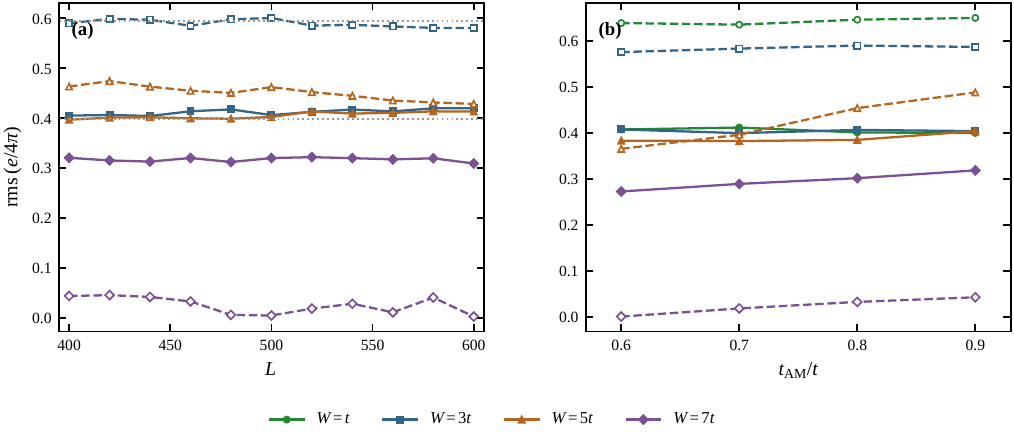}
        \caption{\label{fig:robustness}
        Robustness of the two plateaus at $E=t$ and $\theta=\pi/4$. (a) System-size dependence at $t_{\mathrm{AM}}=0.9t$ for $W=3t$, $5t$, and $7t$, for the longitudinal (open symbols) and the spin-Hall (filled symbols) fluctuations. (b) Altermagnetic-strength dependence at $L=400$ for $W=t$, $3t$, $5t$, and $7t$, with the same symbol convention. The horizontal dashed lines denote $0.595\,e/(4\pi)$ and $0.399\,e/(4\pi)$ for the longitudinal and spin-Hall responses, respectively. Each point is calculated from at least 1000 disorder realizations.
        }
    \end{figure*}

    The disorder scans in the Letter show broad rms plateaus near $0.60\,e/(4\pi)$ and $0.40\,e/(4\pi)$ at $\theta=\pi/4$ for the longitudinal and the spin-Hall response, respectively. We now examine the dependence of both on sample size and altermagnetic strength. Figure~\ref{fig:robustness} summarizes these additional scans, all performed at $E=t$ and $\theta=\pi/4$ with at least 1000 disorder realizations per point.

    Figure~\ref{fig:robustness}(a) shows the size dependence at $t_{\mathrm{AM}}=0.9t$. At $W=3t$, $\mathrm{rms}(\sigma_{\mathrm{s}})$ stays between $0.571$ and $0.604\,e/(4\pi)$ and $\mathrm{rms}(\sigma_{\mathrm{sH}})$ between $0.396$ and $0.421\,e/(4\pi)$ as $L$ is varied from $400$ to $600$, with no systematic trend in either case. At $W=5t$ the longitudinal amplitude decreases from $0.46$ to $0.43\,e/(4\pi)$ over the same range, whereas the spin-Hall value stays within $0.40$--$0.42\,e/(4\pi)$. The $W=7t$ longitudinal amplitude collapses below $0.05\,e/(4\pi)$, while the spin-Hall value stays near $0.31\,e/(4\pi)$; both lie below their plateau values, consistent with the strong-disorder suppression seen in the main-text disorder scans. 
    The two intermediate-disorder plateaus therefore persist over the tested range of system sizes. This supports finite-size robustness over that range, without establishing the thermodynamic limit.
    
    Figure~\ref{fig:robustness}(b) shows the corresponding altermagnetic-strength scan at $L=400$. For $0.6t\leq t_{\mathrm{AM}}\leq0.9t$, the $W=t$ and $3t$ curves remain within $0.400$--$0.412\,e/(4\pi)$ for the spin-Hall response, whereas the longitudinal values lie within $0.635$--$0.650\,e/(4\pi)$ for $W=t$ and $0.576$--$0.590\,e/(4\pi)$ for $W=3t$. The $W=5t$ curves approach these scales as $t_{\mathrm{AM}}$ increases over the same interval. Outside the tested plateau region, strong disorder or weak altermagnetic splitting can shift the rms away from these values. The near-parameter-independence discussed in the Letter applies within the plateau windows and does not extend across the full parameter space.

\section{Orientation-resolved disorder crossovers}\label{app:angles}

    The angular scans in Fig.~\ref{fig:sm_fluctuations}(b,e) show that both spin responses vary strongly with the crystalline orientation at fixed disorder. 
    To compare the disorder evolution at fixed orientations, we examine full disorder scans at $\theta=0$, $\pi/12$, $\pi/6$, and $\pi/4$.
    The calculations use $E=t$, $L=400$, and $t_{\mathrm{AM}}=0.9t$, matching Fig.~3(a,c) of the Letter. Each parameter point uses an ensemble of at least 1000 disorder realizations.
    
    \begin{figure}[t]
        \centering
        \includegraphics[width=0.72\textwidth]{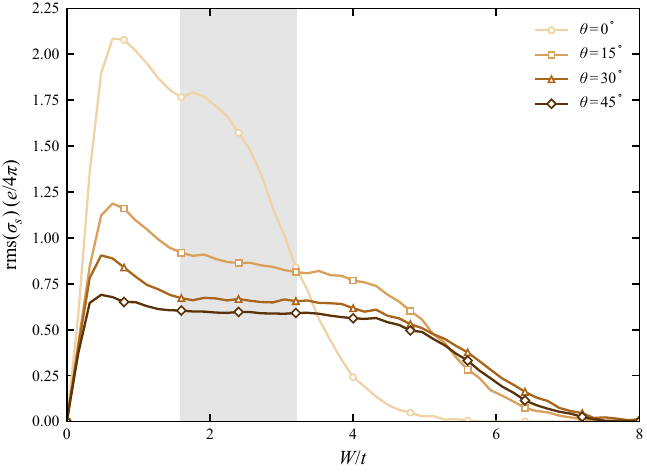}
        \caption{\label{fig:angle_plateaus}
        Disorder dependence of the two-terminal longitudinal spin-conductance fluctuation at four crystalline orientations. 
        The gray band marks the common comparison interval $1.6\leq W/t\leq3.2$.
        Each curve rises from the clean limit, develops a plateau-like region with an orientation-dependent residual slope, and decreases at stronger disorder.
        The calculations use $E=t$, $L=400$, and $t_{\mathrm{AM}}=0.9t$, with at least 1000 disorder realizations at every point. The $\theta=\pi/4$ curve is the corresponding dataset shown in Fig.~\ref{fig:sm_fluctuations}(c).}
    \end{figure}
    
    As shown in Fig.~\ref{fig:angle_plateaus}, the four orientations exhibit plateau-like regions with different amplitudes and residual slopes before the strong-disorder decrease.
    Averaging over the shaded interval gives $1.465\,e/(4\pi)$ at $\theta=0$ and $0.868$, $0.662$, and $0.598\,e/(4\pi)$ at $\theta=\pi/12$, $\pi/6$, and $\pi/4$, respectively. The corresponding window spreads are $0.333$, $0.034$, $0.008$, and $0.008$, in units of $e/(4\pi)$. 
    The points near $W=t,3t$ and $5t$ are consistent with the angular dependence in the Letter.
    The distinct amplitudes and residual slopes persist across the common comparison window, showing that orientation changes both the fluctuation scale and its disorder dependence.

\begin{figure*}[t]
        \centering
        \includegraphics[width=0.96\textwidth]{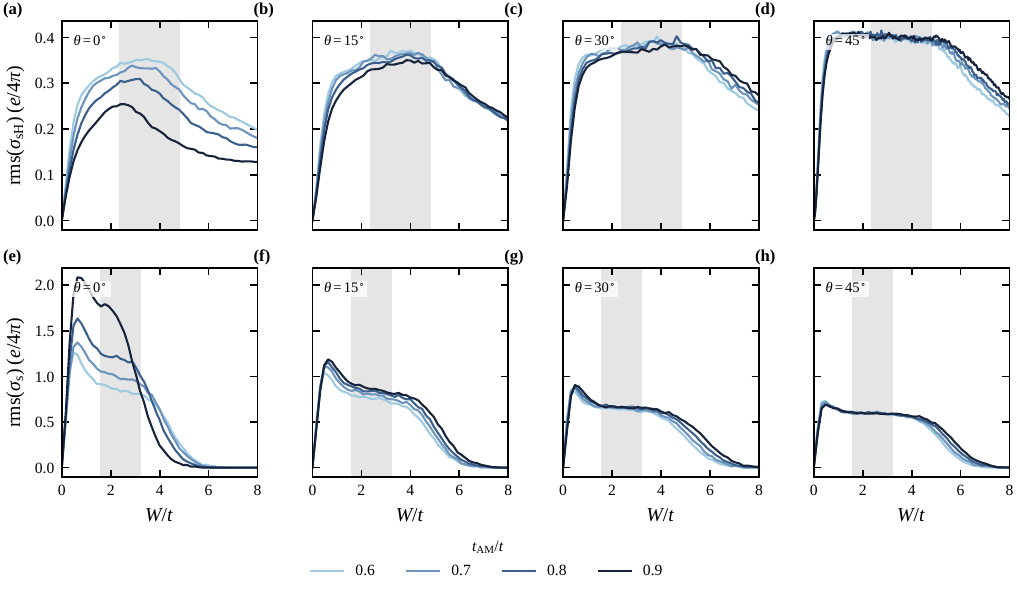}
        \caption{\label{fig:spin_hall_angle_plateaus}
        Disorder dependence of the spin-Hall (a--d) and longitudinal (e--h) conductance fluctuations at $\theta=0$, $\pi/12$, $\pi/6$, and $\pi/4$. Each panel contains $t_{\mathrm{AM}}=0.6t$, $0.7t$, $0.8t$, and $0.9t$. The gray bands mark the comparison windows $2.4\leq W/t\leq4.8$ for the spin-Hall and $1.6\leq W/t\leq3.2$ for the longitudinal response; the corresponding window averages are quoted in the text. The curves become flatter and less dependent on $t_{\mathrm{AM}}$ as $\theta$ approaches $\pi/4$. All calculations use $E=t$, $L=400$, and at least 1000 disorder realizations per point.}
    \end{figure*}

    We perform four-terminal spin-Hall disorder scans at the same four orientations for $t_{\mathrm{AM}}=0.6t$--$0.9t$. The $0.9t$ curves at these angles appear in Fig.~3(c) of the Letter, and the $\pi/4$ strength comparison appears in Fig.~3(d). Figure~\ref{fig:spin_hall_angle_plateaus} shows flatter plateaus and weaker strength dependence as the sampled orientations approach $\theta=\pi/4$. 
    We average over $2.4\leq W/t\leq4.8$ and the four altermagnetic strengths at each orientation. The mean amplitudes are $0.291\,e/(4\pi)$ at $\theta=0$ and $0.353\,e/(4\pi)$, $0.381\,e/(4\pi)$, and $0.399\,e/(4\pi)$ at $\theta=\pi/12$, $\pi/6$, and $\pi/4$, respectively. The corresponding spreads are $0.054$, $0.010$, $0.009$, and $0.006$, in units of $e/(4\pi)$. 
As defined in Sec.~\ref{app:ensemble}, these spreads include residual parameter dependence and finite-ensemble variability.

    The same construction applied to the longitudinal response gives strength scans at the four orientations. Averaging over $1.6\leq W/t\leq3.2$ and the four strengths yields $1.120$, $0.817$, $0.654$, and $0.595\,e/(4\pi)$ at $\theta=0$, $\pi/12$, $\pi/6$, and $\pi/4$, with spreads $0.286$, $0.051$, $0.017$, and $0.009$. Both responses therefore develop flatter plateaus and weaker dependence on $t_{\mathrm{AM}}$ as the orientation approaches $\pi/4$: between $\theta=0$ and $\pi/4$ the longitudinal spread falls by a factor of about $32$ and the spin-Hall spread by a factor of about $9$.

\section{Fermi-velocity anisotropy}\label{app:velocity}

    \begin{figure*}[t]
        \centering
        \includegraphics[width=0.88\textwidth]{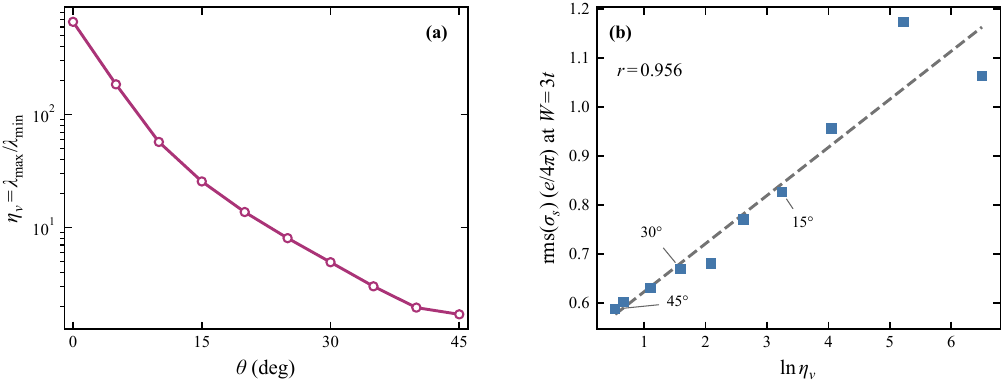}
        \caption{\label{fig:velocity_anisotropy}
        Microscopic measure of transport anisotropy at $E_{\mathrm{F}}=t$ and $t_{\mathrm{AM}}=0.9t$. (a) Eigenvalue ratio $\eta_{\mathrm{v}}$ of the spin-resolved Fermi-surface velocity tensor defined in Eq.~\eqref{eq:velocity_tensor}. (b) Longitudinal spin-conductance fluctuation at $W=3t$ versus $\ln\eta_{\mathrm{v}}$ over $0\leq\theta\leq\pi/4$ in $5^\circ$ increments. The dashed line is a linear guide to the eye; the Pearson correlation coefficient is $r=0.956$.}
    \end{figure*}

    To characterize the clean-band anisotropy associated with the longitudinal fluctuations, we evaluate the spin-resolved Fermi-surface velocity tensor
    \begin{equation}
        \begin{gathered}
        V_{ij}^{(s)}=
        \frac{\displaystyle\int_{\mathrm{BZ}}\md^2k\,
        v_{i,s}(\bm{k})v_{j,s}(\bm{k})
        \delta_{\gamma}\!\left[E_s(\bm{k})-E_{\mathrm{F}}\right]}
        {\displaystyle\int_{\mathrm{BZ}}\md^2k\,
        \delta_{\gamma}\!\left[E_s(\bm{k})-E_{\mathrm{F}}\right]},
        \\
        \delta_{\gamma}(x)=\frac{\gamma}{\pi(x^2+\gamma^2)},
        \end{gathered}
        \label{eq:velocity_tensor}
    \end{equation}
    where $v_{i,s}=\hbar^{-1}\partial E_{s}/\partial k_i$, with $a=\hbar=1$ in the numerical evaluation, $E_{\mathrm{F}}=t,t_{\mathrm{AM}}=0.9t$, and $\gamma = 0.032t$.
    We characterize the anisotropy by the coordinate-independent eigenvalue ratio $\eta_{\mathrm{v}}=\lambda_{\max}(V^{(s)})/\lambda_{\min}(V^{(s)})$. 
    The clean dispersions of the two spin blocks are related by a $\pi/2$ rotation.
    Their velocity tensors therefore have the same eigenvalues, so either block gives the same $\eta_{\mathrm{v}}$. 
    For nondegenerate eigenvalues, the principal axes are related by the same rotation.
    
    Within a semiclassical relaxation-time approximation with a momentum-independent transport time, the diffusion tensor is $D_{ij}^{(s)}\simeq\tau_{\mathrm{el}}^{(s)}V_{ij}^{(s)}$.
    This approximation motivates the clean-band anisotropy measure without determining the diffusion tensor quantitatively over the full disorder range.
    The conductance-fluctuation correlator can be expressed as a sum over eigenmodes of the diffusion operator~\cite{LeeStoneFukuyama1987}. For the present two-dimensional device, these modes satisfy the schematic equation
    \begin{equation}
        \left[-\partial_iD_{ij}^{(s)}\partial_j+\tau_{\phi}^{-1}\right]Q_{\nu}^{(s)}
        =\Lambda_{\nu}^{(s)}Q_{\nu}^{(s)}.
        \label{eq:anisotropic_diffusion_modes}
    \end{equation}
    The device boundaries remain fixed as the orientation and eigenvalue ratio of $D_{ij}^{(s)}$ change. Consequently, both the diffusion spectrum $\{\Lambda_{\nu}^{(s)}\}$ and the associated fluctuation mode sum can depend on $\theta$ without a change of Wigner--Dyson class. The quantity $\eta_{\mathrm{v}}$ characterizes the band anisotropy underlying Eq.~\eqref{eq:anisotropic_diffusion_modes}. The correlation analysis below does not solve the full four-terminal boundary-value problem.

    The endpoint $\theta=0$ provides an exact illustration of the hopping anisotropy.
    The spin-resolved dispersion can then be written
    \begin{equation}
        \begin{aligned}
        E_{\sigma}(\bm{k},0)={}&-2(t-\sigma t_{\mathrm{AM}})\cos k_x\\
        &-2(t+\sigma t_{\mathrm{AM}})\cos k_y.
        \end{aligned}
        \label{eq:channel_blocking_limit}
    \end{equation}
    At $t_{\mathrm{AM}}=t$, the $\sigma=+1$ channel has no $k_x$ dispersion, so its group velocity along the transport direction vanishes and that channel is completely blocked. The large but finite $\eta_{\mathrm{v}}$ near $\theta=0$ at $t_{\mathrm{AM}}=0.9t$ is consistent with proximity to this channel-blocking limit [Fig.~\ref{fig:velocity_anisotropy}(a)].

    Figure~\ref{fig:velocity_anisotropy}(a) shows that $\eta_{\mathrm{v}}$ decreases rapidly as $\theta$ approaches $\pi/4$. In particular, $\eta_{\mathrm{v}}=25.6$, $4.92$, and $1.70$ at $\theta=\pi/12$, $\pi/6$, and $\pi/4$. The same ordering occurs in the longitudinal plateau amplitudes $0.868$, $0.662$, and $0.598\,e/(4\pi)$, while the spin-Hall amplitudes run in the opposite direction. Over the irreducible interval $0\leq\theta\leq\pi/4$, sampled every $5^\circ$, the $W=3t$ longitudinal fluctuation and $\ln\eta_{\mathrm{v}}$ have a Pearson correlation coefficient of $0.956$ [Fig.~\ref{fig:velocity_anisotropy}(b)]; the corresponding spin-Hall series, obtained by the same construction, has $r=-0.988$ (Fig.~4(a) of the Letter). These two coefficients link the fluctuation amplitudes to the clean-band anisotropy without providing an analytic expression for either amplitude.

\section{Channel covariance and comparison with earlier spin-transport fluctuation studies}\label{sm:comparison}
The longitudinal spin conductance is the difference of two conserved-spin
channel conductances, as also occurs in magnetic graphene nanoribbons~\cite{Wimmer2008PRL,Bundesmann2013PRB}.
The channel variances and their covariance therefore both enter the total
variance. Exact spin conservation does not imply statistical independence,
because both blocks experience the same scalar onsite potential. Figure~4(b,c)
of the Letter gives the correlation coefficients and covariance insets at
$\theta=\pi/4$. The longitudinal covariance becomes small beyond the
initial weak-disorder crossover, indicating approximate decorrelation rather than statistical independence. The transverse covariance changes from
negative to positive with increasing $W$, respectively enhancing and
suppressing the total variance under the common up-minus-down convention.
These data describe the disorder evolution at the displayed orientation.

A quantum-transport $T$-odd spin-Hall response of altermagnets has already been studied in
multiterminal devices~\cite{Wei2025PRB}. That work motivates the observable,
whereas the Letter concerns its disorder-ensemble rms, angular dependence,
and plateau development. Similarly, band-anisotropy effects on charge UCF
are established~\cite{Xiao2023NanoLett,Yang2025PRBAnisotropicUCF}; the Letter
examines their conserved-spin counterpart and the role of channel covariance.

Disorder-induced marginal metallic behavior with spin-flip scattering has
also been studied~\cite{Li2026PRLDisorder}. The spin-mixing disorder of that
problem differs from the scalar, exactly spin-conserving disorder here.
The finite-size fluctuation plateaus in the Letter are not a claim of a
stable two-dimensional metallic phase in the thermodynamic limit.

\section{Directional hopping weights and limits of the angular interpretation}\label{sm:angular_interpretation}
Figure~1 of the Letter links band anisotropy and shared scalar disorder to schematic plateau formation. Its crossed arrow marks forbidden transitions between the spin sectors of the same sample. The device geometries appear in Fig.~2(a) of the Letter. Near $\Gamma$, the dispersion becomes, with $a=\hbar=1$,
\begin{equation}
 E_s\simeq -4t+t(k_x^2+k_y^2)
 -s t_{\mathrm{AM}}\left[\cos(2\theta)(k_x^2-k_y^2)
                  +2\sin(2\theta)k_xk_y\right].
\end{equation}
The axial spin contrast is controlled by $\cos(2\theta)$, whereas the mixed-direction term is controlled by $\sin(2\theta)$. The former decreases and the latter increases over the orientations $15^\circ$--$45^\circ$. These factors specify the directional hopping weights, but do not imply proportional changes in the rms amplitudes. In particular, a vanishing mean spin conductance does not imply a vanishing disorder variance.

In the quadratic continuum approximation, the two principal kinetic coefficients are $t\pm t_{\mathrm{AM}}$, independent of $\theta$. Consequently, a rigid rotation of this continuum ellipse alone cannot explain the changing eigenvalue ratio of the full-lattice velocity tensor in Sec.~VII. Both the orientation relative to the fixed boundaries and the lattice-induced change of band shape enter the present calculation. The anisotropy discussed here is a band or transport anisotropy; a magnetocrystalline anisotropy energy has not been computed.

The longitudinal and transverse conductances contain different linear combinations of spin-resolved transmission coefficients. Their variances therefore need not have the same angular dependence. Quantum-transport studies of the $T$-odd spin-Hall effect emphasize the role of transverse boundary conditions, even for the mean response~\cite{Wei2025PRB}. Anisotropy-dependent charge UCF and geometric spin-transport correlations provide further context~\cite{Yang2025PRBAnisotropicUCF,Adagideli2010PRL}. These results motivate comparing the two geometries, but do not predict whether the spin-Hall rms increases or decreases with orientation.

For either measurement geometry $g$, the total variance satisfies the exact relation
\begin{equation}
 V_g(\theta,W)=A_g(\theta,W)-2C_g(\theta,W),\qquad
 A_g=\mathrm{var}(Z_{g\uparrow})+\mathrm{var}(Z_{g\downarrow}),
 \quad C_g=\mathrm{cov}(Z_{g\uparrow},Z_{g\downarrow}).
\end{equation}
The data show opposite angular trends in the longitudinal and spin-Hall rms amplitudes, together with a correlation between longitudinal rms and clean-band velocity anisotropy. The covariance scans at $45^\circ$ resolve disorder dependence but do not separate the angular dependences of $A_g$ and $C_g$. They therefore cannot identify which contribution drives the angular increase in spin-Hall rms.

\FloatBarrier
\bibliographystyle{apsrev4-2}
\bibliography{references}